\documentclass[11pt, a4paper]{article}

\usepackage[margin=2.5cm]{geometry} 
\usepackage{setspace}
\usepackage[utf8]{inputenc}
\usepackage[T1]{fontenc}

\usepackage{amsmath, amsthm, amssymb} 
\usepackage{mathtools}

\usepackage{titlesec}
\titleformat*{\section}{\large\bfseries}
\titleformat*{\subsection}{\normalsize\bfseries}

\titleformat{\paragraph}[runin]
  {\normalfont\slshape}
  {\theparagraph}
  {1em}
  {}
  [. ]

\newcommand{\E}{\mathbb{E}}

\newcommand{\sym}[1]{\rlap{$^{#1}$}}

\usepackage{caption}
\usepackage{subcaption}
\usepackage{booktabs}
\usepackage{threeparttable}
\usepackage{siunitx}
\usepackage{multirow}
\usepackage{tabularx}
\usepackage{etoolbox}
\robustify\bfseries

\usepackage[round,authoryear]{natbib}
\usepackage[colorlinks=true,allcolors=blue]{hyperref}

\usepackage{authblk}

\theoremstyle{plain}

\theoremstyle{definition}

\title{\large{\textbf{
Structural Cointegration of the Climate--Carbon Feedback: \\Evidence from the Last 130,000 Years}
}}

\author[1]{\text{Satoshi Nakano}\thanks{Email: nakano@n-fukushi.ac.jp}}
\author[2]{\text{Kazuhiko Nishimura}\thanks{Email: nishimura@lets.chukyo-u.ac.jp}}

\affil[1]{Faculty of Economics, Nihon Fukushi University, Japan}
\affil[2]{Institute of Economics, Chukyo University, Japan}

\date{\small \today} 

\begin{document}
\maketitle
\begin{abstract}
\noindent
Using a gap-free, millennial-resolution ice-core record spanning the last 130,000 years, we identify the feedback architecture between Antarctic temperature and atmospheric CO$_2$. The series are found to be cointegrated, justifying estimation with a Vector Error Correction Model (VECM). The estimated long-run relationship yields a temperature change of 13.0 K per CO$_2$ doubling. Structural identification combining Milankovitch-cycle instrumental variables with the VECM residuals yields a contemporaneous carbon response (CCR) that closely aligns with the thermodynamic bounds of Henry's Law.
For the contemporaneous temperature response (CTR), the estimated confidence interval fully encompasses the theoretical Planck response. 
Accounting for structural lags and error correction, the cumulative temperature response within one millennium of CO$_2$ doubling reaches 11.8 K, successfully quantifying the delayed manifestation of the greenhouse effect.
\end{abstract}
\section{Introduction} 
The tight coupling between Antarctic temperature and atmospheric carbon dioxide (CO$_2$) concentrations over the late Quaternary is one of the most striking features of the Earth system. 
While the high degree of correlation between these variables is well-established, the nature of their contemporaneous interaction---specifically the direction, magnitude, and timing of feedbacks at millennial resolutions---remains a subject of intense debate.
Establishing a robust causal architecture is not only vital for understanding past glacial-interglacial transitions but also for constraining long-term Earth System Sensitivity (ESS) in the context of anthropogenic warming.

The debate over the phase relationship between these variables has been long-standing. 
\citet{Caillon2003} famously reported that Antarctic temperature leads CO$_2$ by approximately 800 years during Termination III, a finding often used to argue that the initial climate shift is driven by orbital forcing rather than greenhouse gases. 
However, as \citet{Lemoine2010} pointed out, such simple lead-lag analyses based on peak-to-peak timing fail to disentangle the mutual causal effects from autocorrelation and reverse causality. 
In a complex system where temperature and carbon are mutually endogenous, a lead in one variable does not preclude the existence of a powerful, simultaneous feedback from the other. 
Without a framework that accounts for this reciprocity, the magnitude of the positive climate--carbon feedback may be significantly underestimated.

Furthermore, the statistical analysis of paleoclimate records faces fundamental analytical challenges. 
A primary concern is the high persistence of temperature and CO$_2$ series, which carries the risk of \textsl{spurious regression} \citep{Kaufmann2006, Cummins2022}. 
Because these variables exhibit non-stationary, $I(1)$ properties, a cointegration framework is essential to separate true physical bonds from stochastic trends. 
In this context, \citet{Kaufmann2013} provided a seminal benchmark by estimating a Cointegrated Vector Autoregression (CVAR) model; however, as noted by \citet{Miller2019}, such models often rely on interpolated data, which can introduce significant statistical artifacts. 
Furthermore, even when multivariate models are employed to capture slow internal dynamics \citep{Kaufmann2016}, the identification of the contemporaneous interaction matrix ($A_0$) has historically relied on pre-assumed physical zero-constraints. 
Consequently, the objective allocation of causality within a single millennial time step---the very scale where the initial feedback triggers occur---remains an unresolved empirical question.

In this paper, we address these challenges through a unified empirical strategy. 
First, we focus our primary structural analysis on a specific 130,000-year interval since the Last Interglacial that is entirely free of observational gaps at a 1,000-year resolution. By utilizing this gap-free sample, we eliminate the need for interpolation, thereby neutralizing the potential artifacts warned of by \citet{Miller2019}. 
Second, we employ a bivariate framework (temperature and CO$_2$) rather than expanding the model to include highly collinear variables such as ice volume. Because ice volume variations are tightly coupled with global temperature, forcing both into a simultaneous identification scheme risks severe multicollinearity. By treating temperature as a macroscopic sufficient statistic, our reduced-form VECM objectively captures the comprehensive \textsl{total effect} of the climate-carbon interaction, seamlessly internalizing the indirect forcing from slower cryospheric feedbacks. 
Third, we utilize the strict exogeneity of Milankovitch cycles as an instrumental variable (IV). This allows us to statistically disentangle the contemporaneous temperature response (CTR, $a_{12}$) from the contemporaneous carbon response (CCR, $a_{21}$) within the VECM residuals without relying on ad-hoc zero-constraints.


Our results reveal a stark asymmetry in the contemporaneous climate--carbon interaction that perfectly bridges time-series econometrics with thermodynamic constraints. We find a robust, statistically significant CCR ($\hat{a}_{21} > 0$), strictly consistent with the theoretical boundaries of Henry's Law for oceanic outgassing. Regarding the CTR ($a_{12}$), its estimated 95\% confidence interval fully encompasses the theoretical Planck response. This confirms that while external temperature changes trigger an immediate carbon release, the instantaneous radiative feedback of CO$_2$ onto temperature is severely buffered by the massive thermal inertia of the oceans and cryosphere, requiring structural lags to fully manifest. Ultimately, we provide a unified physical explanation for the perceived ``synchronicity'' in deglacial records as a manifestation of the immediate CCR, while reconciling the apparent ``lag'' as the structural adjustment time required to overcome thermal debt.

\section{Model}

In this section, we delineate the mathematical and physical foundations of our empirical approach. Given the high degree of persistence and the inherent feedbacks between temperature and greenhouse gas concentrations observed in ice-core records, a robust identification strategy must account for both long-term equilibrium states and short-term structural innovations.\footnote{For a more exhaustive mathematical background on the Vector Autoregressive (VAR) and Vector Error Correction (VEC) frameworks, including the properties of non-stationary systems and the derivation of the long-run permanent component ($\Xi$), the reader is referred to Appendix \ref{apd2}.}

\subsection{Vector Error Correction Model}

A standard Vector Autoregressive (VAR) model typically assumes that the underlying time series are stationary. 
However, paleoclimatic proxies such as temperature and CO$_2$ concentrations exhibit the properties of non-stationary, $I(1)$ processes. 
Estimating a VAR model in levels for such variables carries the risk of \textit{spurious regression}, where the high degree of persistence leads to statistically significant correlations that do not represent true physical causalities.

To circumvent this risk while preserving the critical information contained in the levels of the variables, we employ a Vector Error Correction Model (VECM). 
This framework is appropriate when variables are \textit{cointegrated}, meaning that while individual series may drift non-stationarily, a stable linear combination exists. 
In this study, we consider a three-variable system $y_t = (T_t, \ln C_t, M_t)^\prime$, where $T_t$ is Antarctic temperature, $\ln C_t$ is the logarithm of CO$_2$ concentration, and $M_t$ represents external orbital (Milankovitch) forcing. 



We begin with a Structural Vector Autoregressive (SVAR) model that explicitly captures contemporaneous interactions via the matrix $A_0$:
\begin{equation}
A_0 y_t = A_1 y_{t-1} + \dots + A_p y_{t-p} + u_t,
\end{equation}
where $u_t \sim i.i.d. \mathcal{N}(0, \Sigma_u)$ represents the structural innovations. The corresponding reduced-form representation is given by multiplying both sides by $A_0^{-1}$, yielding $y_t = \Phi_1 y_{t-1} + \dots + \Phi_p y_{t-p} + \epsilon_t$, as detailed in Equation \eqref{VARp2} in Appendix \ref{apd2}. Comparing the two representations, the reduced-form residuals are linked to the structural shocks by the relationship $A_0 \epsilon_t = u_t$.

Moreover, the reduced-form VAR \eqref{VARp2} can be reformulated as a VECM:
\begin{equation}
\Delta y_t = \alpha \beta^\prime y_{t-1} + \Gamma_1 \Delta y_{t-1} + \dots + \Gamma_{p-1} \Delta y_{t-p+1} + \epsilon_t,
\label{eq:vecm}
\end{equation}
where $\beta$ contains the cointegrating vector, $\alpha$ represents the adjustment speeds toward equilibrium, and $\Gamma_i$ captures the short-run lagged dynamics.
The error term $\epsilon_t$ denotes the reduced-form innovations, which are generally correlated element-wise.

\subsection{Structural Identification and Causality Allocation}

To isolate structural shocks ($u_t$) from the reduced-form residuals ($\epsilon_t$), we utilize the standard structural relationship $A_0 \epsilon_t = u_t$. 
This approach follows the framework established by \citet{Sims1980}, which provides a basis for analyzing endogeneity and structural innovations within a system of interrelated time series by utilizing the residuals of the estimated model.
In our three-variable system, the contemporaneous interaction matrix $A_0$ is defined with ones on the principal diagonal to normalize the structural shocks:
\begin{equation*}
A_0 = \begin{pmatrix} 1 & -a_{12} & -a_{13} \\ -a_{21} & 1 & -a_{23} \\ 0 & 0 & 1 \end{pmatrix} 
\end{equation*}
where $a_{12}$ and $a_{21}$ represent the contemporaneous temperature response (CTR) and the contemporaneous carbon response (CCR), respectively. 
The off-diagonal elements in the third row are set to zero due to the strict exogeneity of orbital forcing, implying $\epsilon_{3t} = u_{3t}$. The structural equations for the climate residuals are then given by:
\begin{align*}
u_{1t} &= \epsilon_{1t} - a_{12} \epsilon_{2t} - a_{13} \epsilon_{3t} 
\\
u_{2t} &= -a_{21} \epsilon_{1t} + \epsilon_{2t} - a_{23} \epsilon_{3t} 
\end{align*}

By imposing the orthogonality condition $\mathbb{E}[u_{1t} u_{2t}] = 0$ and noting that $\epsilon_{3t}$ is orthogonal to both $u_{1t}$ and $u_{2t}$ (since $\epsilon_{3t} = u_{3t}$), the terms involving orbital forcing effectively vanish in the expectation. This yields the Causality Allocation Curve (CAC), which defines the structural trade-off between the CTR and the CCR:
\begin{equation}
(1 + a_{12} a_{21}) s_{12} - a_{21} s_{11} - a_{12} s_{22} = 0 \label{eq:cac}
\end{equation}
where $s_{ij}$ denotes the sample covariance between $\epsilon_{it}$ and $\epsilon_{jt}$.
Equation \eqref{eq:cac} demonstrates that the allocation of the total contemporaneous correlation between temperature and CO$_2$ into directional causal links is uniquely determined by the sample covariance structure ($s_{ij}$) of the residuals.

\subsection{Impulse Response Analysis}

To visualize the propagation of these structural innovations, we utilize Impulse Response Functions (IRFs). The IRFs trace the time path of the system in response to a one-time shock in $u_t$. We simulate two primary scenarios:
\begin{enumerate}
    \item \textsl{A CO$_2$ doubling shock:} Represented as $u_{2t} = \ln(2) \approx 0.693$. This allows us to observe the transition from the transient response to the long-term Earth System Sensitivity (ESS) determined by $\beta$.
    \item \textsl{A 2 K temperature shock:} Represented as $u_{1t} = 2.0$. This isolates the magnitude and speed of the carbon cycle's response via the CCR and the subsequent feedback loops.
\end{enumerate}

The presence of cointegration ensures that temporary structural shocks lead to a permanent shift in the climatic baseline. The long-run impact matrix, defined as $\Xi = \lim_{h \to \infty} \Psi_h$ (see Appendix \ref{apd2} for details), provides a formal bridge between millennial-scale innovations and the long-term equilibrium states.
\section{Data}

\subsection{Sources}

\begin{figure}[t!]
\centering
\includegraphics[width=1.0\textwidth]{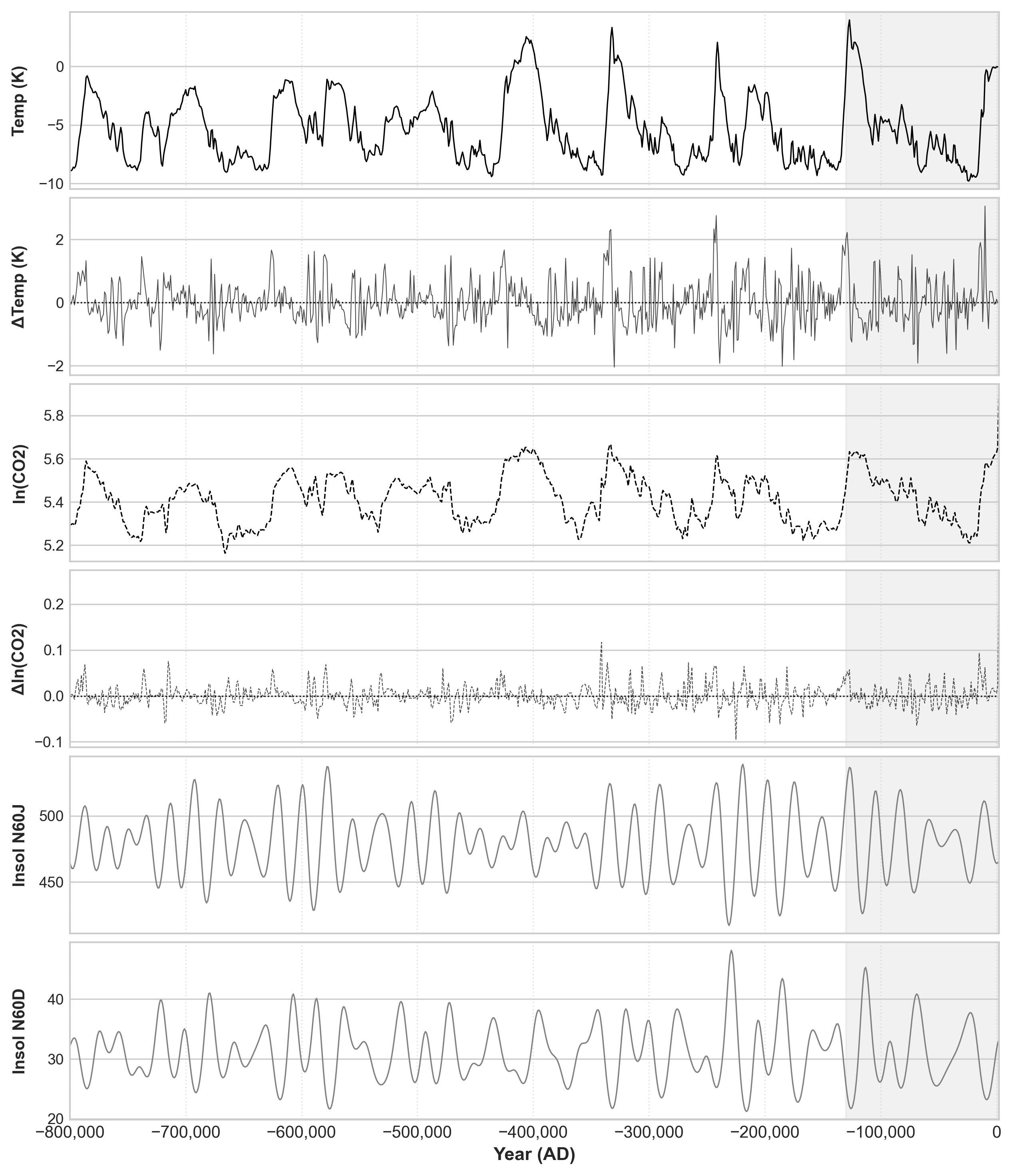}
\caption{Evolution of climatic and orbital variables over the past 800,000 years. The panels display (from top to bottom) solar insolation at $60^\circ$N in June and December, Antarctic temperature anomalies ($\Delta T$), and the logarithm of atmospheric CO$_2$ concentration ($\ln C$). The shaded area indicates the 130 ka sample period used for primary structural analysis.}
\label{fig:series}
\end{figure}

The empirical foundation of this study rests on three primary data series spanning the late Quaternary: atmospheric CO$_2$ concentrations, Antarctic surface temperature anomalies, and solar insolation. Figure \ref{fig:series} illustrates the evolution of these variables over the past 800,000 years, highlighting the strong phase-locking of the climate--carbon system to orbital cycles.

Atmospheric CO$_2$ data are obtained from the composite ice-core record \citep{Luthi2008, Bereiter2015}, which integrates measurements from multiple Antarctic cores to capture natural greenhouse gas variability across eight complete glacial-interglacial cycles. 
For temperature estimates, we utilize the Antarctic temperature anomaly record ($\Delta T$) derived from deuterium ratios in the EPICA Dome C ice core \citep{Jouzel2007}, serving as a reliable indicator for global climatic shifts. 
External orbital forcing is represented by the solar insolation at 60$^\circ$N in June, computed based on the solution by \citet{Laskar2004}.

\subsection{Gridding and Sample Selection}

To conduct a rigorous time-series analysis, the unevenly spaced raw data were transformed into a regular 1,000-year resolution grid. 
We distinguish between two primary sample periods: the comprehensive \textsl{800 ka sample} and the focused \textsl{130 ka sample} (covering the period from the Last Interglacial to the present).

\begin{figure}[t!]
\centering
\includegraphics[width=1.0\textwidth]{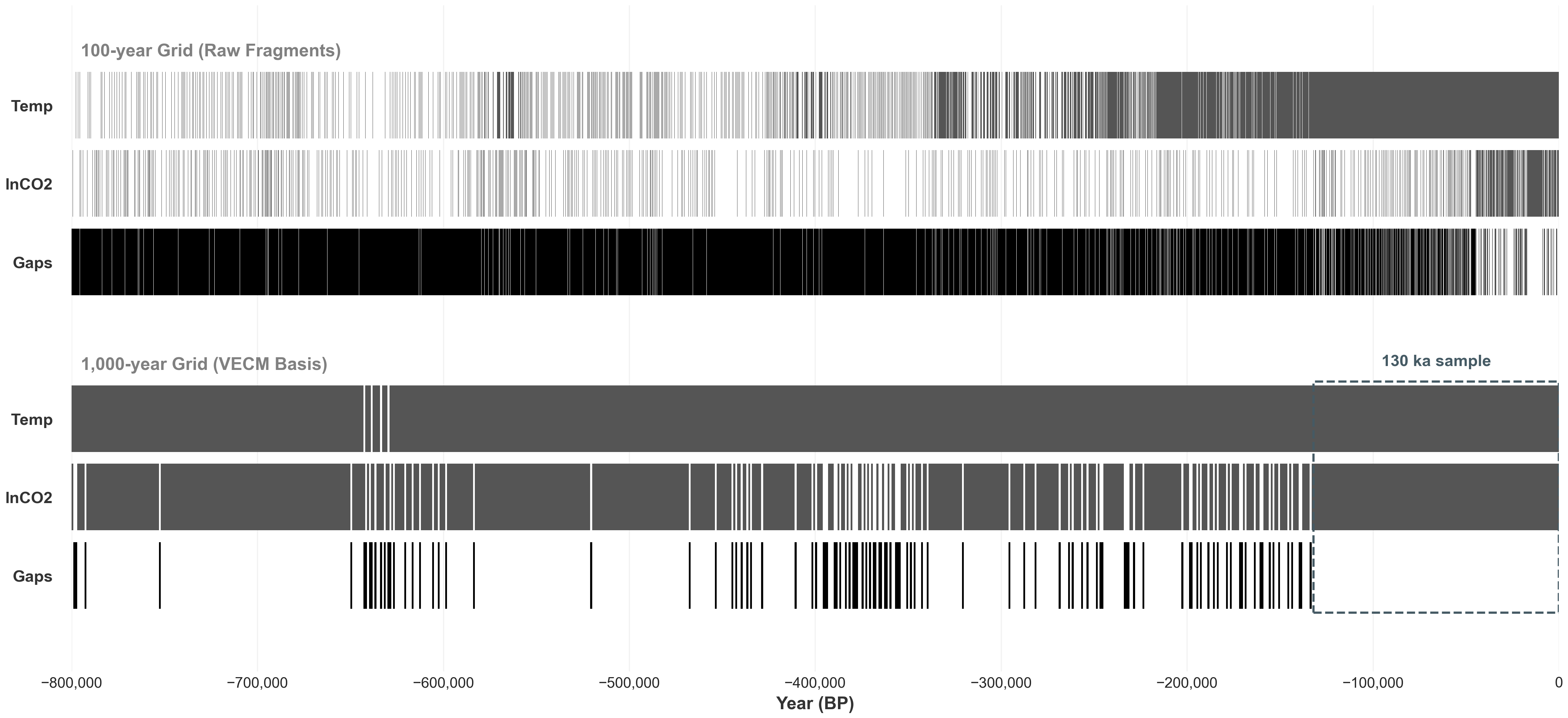}

\caption{Visualization of data gridding and observational gaps. 
The upper panels show the 100-year raw fragments, while the lower panels display the 1,000-year grid used as the VECM basis. 
The dashed box highlights the 130 ka sample as the longest continuous interval free of gaps for both temperature and CO$_2$, justifying its use for structural identification.}
\label{fig:grid}
\end{figure}

Critically, as visualized in Figure \ref{fig:grid}, our grid analysis identifies that the 130 ka sample represents a continuous interval entirely free of observational gaps at the millennial resolution. 
Because this interval requires zero interpolation, the resulting series remains a direct reflection of the physical ice-core record, immune to statistical artifacts.\footnote{For a concrete demonstration of how data interpolation introduces severe structural distortions and causes the causality allocation to collapse, see the supplementary estimation using the interpolated 800 ka record in Appendix \ref{apd1}.} 
By focusing on this gap-free 130 ka record, we ensure that the estimated structural parameters---particularly the CCR ($a_{21}$) and structural lag effects---reflect pure physical signals.

\section{Analysis}
\subsection{VECM Estimation Results}

\begin{table}[t!]
\centering
\caption{Unit Root and Cointegration Test Results}
\label{tab:DF_EG}
\vspace{-1mm}
\begin{tabular*}{\textwidth}{@{\extracolsep{\fill}} l *{4}{S[table-format=-1.3, table-space-text-post={***}]} }
\toprule
 & \multicolumn{2}{c}{\text{130 ka sample}} & \multicolumn{2}{c}{\text{800 ka sample}} \\
\cmidrule(lr){2-3} \cmidrule(lr){4-5}
\textsl{ADF Unit Root Test} & {$Z(t)$} & {$p$-value} & {$Z(t)$} & {$p$-value} \\
\midrule
Temperature ($T$)     & -1.227 & 0.662 & -3.303 & 0.015 \\
$\ln$CO$_2$ ($\ln C$) & -0.565 & 0.879 & -1.635 & 0.465 \\
\midrule
\textsl{Engle-Granger Test} & {$Z(t)$} & {10\% Crit.} & {$Z(t)$} & {1\% Crit.} \\
\midrule
Cointegration ($T$ on $\ln C$) & -3.176\sym{*} & -3.077 & -7.871\sym{***} & -3.910 \\
\bottomrule
\end{tabular*}
\smallskip
\begin{minipage}{\textwidth}
    \footnotesize \textit{Notes:} 
    ADF tests are performed with a constant. Critical values for the Engle-Granger test are based on MacKinnon (2010). $^{***}$ and $^{*}$ denote statistical significance at the 1\% and 10\% levels, respectively.
\end{minipage}
\end{table}
\begin{table}[t!]
\centering
\caption{Lag Order Selection Criteria}
\label{tab:VARSOC}
\vspace{-1mm} 
\begin{tabular*}{\textwidth}{@{\extracolsep{\fill}} l *{8}{S[table-format=-1.4, table-space-text-post=*]}}
\toprule
 & \multicolumn{4}{c}{\text{130 ka sample}} & \multicolumn{4}{c}{\text{800 ka sample}} \\
\cmidrule(lr){2-5} \cmidrule(lr){6-9}
Lag & {FPE} & {AIC} & {HQIC} & {SBIC} & {FPE} & {AIC} & {HQIC} & {SBIC} \\
\midrule
0 & .0219    &  1.855    &  1.874    &  1.900    & .0222    &  1.868    &  1.873    &  1.880    \\
1 & .0002    & -2.829    & -2.775    & -2.695    & .0002    & -2.772    & -2.759    & -2.737    \\
2 & .0002    & -3.047    & -2.957    & -2.824    & .0001    & -3.301    & -3.278    & -3.242    \\
3 & .0001\sym{*} & -3.157\sym{*} & -3.030\sym{*} & -2.845\sym{*} & .0001    & -3.375    & -3.343\sym{*} & -3.292\sym{*} \\
4 & .0002    & -3.108    & -2.945    & -2.707    & .0001\sym{*} & -3.375\sym{*} & -3.334    & -3.269    \\
\bottomrule
\end{tabular*}
\smallskip
\begin{minipage}{\textwidth}
    \footnotesize \textit{Notes:} {*} indicates lag order selected by the criterion. AIC: Akaike info criterion (FPE: final prediction error); SBIC: Schwarz Bayesian info criterion; HQIC: Hannan-Quinn info criterion.
\end{minipage}
\end{table}
\begin{table}[t!]
\centering
\caption{Comparison of VECM Results: 130 ka vs. 800 ka samples}
\label{tab:VECM}
\vspace{-1mm}
\begin{tabular*}{\textwidth}{@{\extracolsep{\fill}} 
  l 
  *{4}{S[
    table-format=-3.4, 
    table-space-text-post=\sym{***}, 
    input-symbols={()}, 
    table-align-text-post=false
  ]}
}
\toprule
 & \multicolumn{2}{c}{\text{130 ka sample}} & \multicolumn{2}{c}{\text{800 ka sample}} \\
\cmidrule(lr){2-3} \cmidrule(lr){4-5}
\multicolumn{5}{l}{\textsl{Panel A: Cointegrating Equation ($\beta$)}} \\
\midrule
{Variable} & 
{Coefficient} & {(s.e.)}
& 
{Coefficient} & {(s.e.)}
\\
\midrule
$T_{t-1}$     & 1.000 & {---} & 1.000 & {---} \\
$\ln C_{t-1}$ & -18.748\sym{***} & (2.312) & -20.251\sym{***} & (1.081) \\
$M_{t-1}$     & 0.009 & (0.038) & -0.105\sym{***} & (0.017) \\
Constant      & 103.300 & {---} & 164.974 & {---} \\
\midrule
\multicolumn{5}{l}{\textsl{Panel B: Short-run Dynamics (Error Correction Model)}} \\
\midrule
{Independent Variables} & {$\Delta T_t$} & {$\Delta \ln C_t$} & {$\Delta T_t$} & {$\Delta \ln C_t$} \\
\midrule
Error Correction ($\alpha$) & -0.103\sym{**} & 0.002 & -0.164\sym{***} & -0.0002 \\
                            & (0.042)    & (0.001) & (0.017)      & (0.001) \\
$\Delta T_{t-1}$            & 0.030   & 0.003 & 0.346\sym{***} & 0.003\sym{***} \\
                            & (0.096) & (0.003) & (0.034)     & (0.001) \\
$\Delta T_{t-2}$            & -0.281\sym{***} & 0.0004 & -0.123\sym{***} & 0.003\sym{**} \\
                            & (0.093)     & (0.003)  & (0.033)      & (0.001) \\
$\Delta \ln C_{t-1}$        & 15.201\sym{***} & 0.447\sym{***} & 6.025\sym{***} & 0.600\sym{***} \\
                            & (2.845)      & (0.102)     & (1.033)     & (0.035) \\
$\Delta \ln C_{t-2}$        & -0.514  & -0.307\sym{***} & 0.723   & -0.312\sym{***} \\
                            & (3.008) & (0.108)      & (1.080) & (0.037) \\
$\Delta M_{t-1}$            & -0.093\sym{***} & -0.0005 & 0.146\sym{***} & 0.0000 \\
                            & (0.033)      & (0.001)  & (0.020)     & (0.001) \\
\midrule
Observations & \multicolumn{2}{c}{129} & \multicolumn{2}{c}{799} \\
AIC          & \multicolumn{2}{c}{$-$3.498} & \multicolumn{2}{c}{$-$3.599} \\
\bottomrule
\end{tabular*}
\smallskip
\begin{minipage}{\textwidth}
    \footnotesize \textit{Notes:} Standard errors are in parentheses. $T$, $\ln C$, and $M$ denote temperature, log-CO$_2$, and orbital forcing, respectively. $^{***}$, $^{**}$, and $^{*}$ denote significance at the 1\%, 5\%, and 10\% levels.
\end{minipage}
\end{table}

Prior to estimating the structural dynamics of the climate system, we conduct a series of diagnostic tests to ensure the statistical validity of the Vector Error Correction framework.
Table \ref{tab:DF_EG} reports the results of the Augmented Dickey-Fuller (ADF) unit root tests and the Engle-Granger cointegration tests.
The ADF statistics indicate that both temperature and the logarithm of CO$_2$ are non-stationary $I(1)$ processes across both the 800 ka and the 130 ka samples.
The Engle-Granger tests confirm the existence of a stable cointegrating relationship, providing the empirical basis for the VECM specification defined in Equation \eqref{eq:vecm}.
Furthermore, the lag order selection criteria presented in Table \ref{tab:VARSOC} suggest that a second-order lag structure (where the level-VAR lag $k=3$ corresponds to VECM lag $p=2$) is optimal for capturing the millennial-scale dynamics while maintaining statistical parsimony.

The primary estimation results for the VECM, comparing the gap-free 130 ka sample with the interpolated 800 ka whole period, are summarized in Table \ref{tab:VECM}.
Panel A presents the estimated cointegrating vector ($\beta$), which defines the long-term equilibrium relationship between temperature and CO$_2$. 
In both samples, the coefficient for $\ln C_{t-1}$ is negative and highly significant ($p < 0.01$), indicating a robust positive long-term correlation between climatic warming and atmospheric carbon concentration.

Panel B describes the short-run dynamics and the error correction mechanism. 
The error correction term ($\alpha$) for the temperature equation is negative and statistically significant in both samples ($-$0.103 for the 130 ka sample), confirming that temperature deviations from the long-term carbon-linked equilibrium are corrected over time. 
In contrast, the error correction term for the $\ln C_t$ equation is statistically indistinguishable from zero, suggesting that atmospheric CO$_2$ acts as the primary stochastic driver of the long-term trend in this system. 
The short-run lagged effects ($\Delta T_{t-i}, \Delta \ln C_{t-i}$) also reveal significant millennial-scale persistence and cross-variable interactions. 
These estimated parameters provide the necessary foundation for the structural causality analysis and sensitivity evaluations performed in the subsequent sections.
\subsection{Causality Allocation Analysis}

The structural parameter $a_{21}$ is defined as a key element of the contemporaneous interaction matrix $A_0$ in our SVAR system, representing the simultaneous feedback within a single 1,000-year time step. 
Specifically, $a_{21}$ quantifies the contemporaneous carbon response (CCR), defined as the response of the logarithm of CO$_2$ concentration to a change in temperature:
\begin{equation*}
a_{21} = \frac{\partial \ln C_t}{\partial T_t}
\end{equation*}
To estimate this parameter from the paleoclimate record, we utilize a regression model formulated in first differences:
\begin{equation}
\Delta \ln C_t = a_{21} \Delta T_t + v_t \label{eq:a21_reg}
\end{equation}
Since temperature and CO$_2$ are characterized as $I(1)$ processes, their first-differenced series are stationary, or $I(0)$. 
This transformation ensures that the estimation is not susceptible to the problem of spurious regression, which often plagues level-based analyses of persistent time series.

However, estimating Equation \eqref{eq:a21_reg} via Ordinary Least Squares (OLS) is potentially problematic due to the inherent endogeneity of the climate-carbon system. 
Given the simultaneous nature of the coupling, a reverse causal link also exists:
\begin{equation*}
\Delta T_t = a_{12} \Delta \ln C_t + w_t
\end{equation*}
where $a_{12}$ represents the contemporaneous temperature response (CTR). 
If a significant CTR actively manifests at the millennial resolution, the OLS estimate for $a_{21}$ would suffer from simultaneity bias, as $\Delta T_t$ would be correlated with the error term $v_t$.

To address this concern and ensure the consistency of our estimates, we first employ Instrumental Variables (IV) estimation. 
Our identification strategy utilizes Milankovitch orbital parameters, specifically the winter solstice insolation at 60$^\circ$N ($M^\text{N60D}_t$), and lagged temperature levels ($T_{t-2}$) as instruments.
These celestial mechanics provide a source of exogenous variation in temperature that is physically independent of the Earth's internal carbon cycle. 
By utilizing these instruments, we isolate the variations in temperature that are driven by external forcing, thereby identifying the pure physical response of atmospheric CO$_2$ to climatic warming.

The estimation results are summarized in Table \ref{tab:a21Estim3}. 
Crucially, the endogeneity test (a robust Durbin-Wu-Hausman test) for the 130 ka sample yields a $p$-value of 0.864. 
This result indicates that we cannot reject the null hypothesis that the temperature change is exogenous with respect to CO$_2$ innovations. In the context of our causality allocation, this provides strong empirical evidence that the immediate radiative forcing of CO$_2$ onto temperature (the CTR) is heavily buffered by the massive thermal inertia of the climate system at the 1,000-year resolution.
Consequently, OLS estimation is not only consistent but also more efficient than IV estimation. 
The resulting OLS estimate is $\hat{a}_{21} = 0.02082$ ($z = 8.04, p < 0.01$), which we adopt as the representative structural parameter for the CCR.
\begin{table}[t!]
\centering
\caption{IV Estimation Results for CCR ($a_{21}$)}
\label{tab:a21Estim3}

\begin{tabular*}{\textwidth}{
@{\extracolsep{\fill}}
>{\raggedright\arraybackslash}p{0.32\textwidth}
S[table-format=1.4]
S[table-format=2.2]
S[table-format=1.4]
S[table-format=1.4]
S[table-format=2.2]
}
\hline 
{Sample Period} &
{Estimate} &
{$z$} &
\multicolumn{2}{c}{95\% Conf. Interval} &
{Sensitivity} \\
\cline{4-5} 
&$a_{21}$&& {Lower} & {Upper} & {(ppm/K)} \\
\hline 

800 ka
& 0.0222 & 5.44 & 0.0142 & 0.0302 & 6.22 \\

130 ka
& 0.0219 & 3.81 & 0.0106 & 0.0332 & 6.13 \\

130 ka (OLS)
& 0.0208 & 8.04 & 0.0157 & 0.0259 & 5.82 \\

\hline 
\textsl{Diagnostics for IV Estimation} & & \multicolumn{1}{c}{1st $F$} & \multicolumn{1}{c}{Overid.} & \multicolumn{1}{c}{Endog.} & \multicolumn{1}{c}{Obs.} \\
\hline 

800 ka
& 
& 28.48
& 0.963
& 0.005
& \multicolumn{1}{c}{801} \\

130 ka
& 
& 11.91
& 0.054
& 0.864
& \multicolumn{1}{c}{130} \\

\hline 
\end{tabular*}

\begin{minipage}{\textwidth}\small
\textit{Notes:} 
The dependent variable is $\Delta \ln C_t$ and the endogenous regressor is $\Delta T_t$ as defined in Equation \eqref{eq:a21_reg}. 
Instruments for 800 ka are $\{M^\text{N60J}_t, M^\text{N60D}_t\}$ and for 130 ka are $\{M^\text{N60D}_t, T_{t-2}\}$. 
Overid. reports the $p$-value for the Hansen J or Sargan tests. 
Endog. reports the $p$-value for the Durbin-Wu-Hausman endogeneity test; the high $p$-value 130 ka (0.864) indicates that the null hypothesis of exogeneity for $\Delta T_t$ cannot be rejected, justifying the use of the OLS estimate ($\hat{a}_{21} = 0.0208$). 
Sensitivity is converted as $a_{21} \times 280$ ppm/K for comparison.
\end{minipage}

\end{table}

To further rigorously test the structural parameters while mutually accounting for the uncertainty in both the CCR ($a_{21}$) and the CTR ($a_{12}$), we perform a joint estimation using the Generalized Method of Moments (GMM). 
This approach incorporates both the regression model in Equation \eqref{eq:a21_reg} and the orthogonality condition derived from the structural shock properties, i.e., the CAC, or Equation \eqref{eq:cac}. 
The moment conditions for this joint estimation are defined by the structural identification framework established in Section 2.2.

The results of the GMM joint estimation are presented in Table \ref{tab:GMM_results}. 
The estimate for $a_{21}$ is 0.0206 ($z = 7.24, p < 0.01$), which is highly consistent with the earlier OLS result. 
In contrast, the estimate for $a_{12}$ is 0.2209 with an exceptionally large standard error of 2.5253. 
Rather than testing whether this parameter is statistically zero, we evaluate it against the fundamental physical baseline: the theoretical Planck response (1.644 K). A Wald test for the null hypothesis that the CTR equals the theoretical Planck response ($H_0: a_{12} = 1.644$ ---which corresponds to an intrinsic warming of 1.14 K per CO$_2$ doubling without feedbacks) yields a $\chi^2(1)$ statistic of 0.32 with a $p$-value of 0.57, meaning we comfortably fail to reject this physically grounded hypothesis. 
This robustly confirms that while the immediate radiative forcing of CO$_2$ is suppressed by the thermal inertia of the oceans and cryosphere, the underlying thermodynamic constraints are strictly maintained.
\begin{table}[t!]
\centering
\caption{GMM Joint Estimation Results and Wald Test for Structural Parameters}
\label{tab:GMM_results}

\begin{tabularx}{\textwidth}{ 
>{\raggedright\arraybackslash}X 
S[table-format=1.4, table-column-width=1.4cm] 
S[table-format=1.4] 
S[table-format=1.4] 
S[table-format=1.4] 
S[table-format=2.2] 
}

\hline
 & {Coef.} & {Std. Err.}
 & \multicolumn{2}{c}{95\% Conf. Interval}
 & {Sensitivity} \\
\cline{4-5}

\text{Parameter}
 &  &  & {Lower} & {Upper} & {(ppm/K)} \\

\hline

$a_{21}$
& {0.0206\sym{***}} & 0.0029
& 0.0151 & 0.0262
& 5.77 \\

$a_{12}$
& 0.2209 & 2.5253
& -4.7286 & 5.1703
&  \\

Constant
& 0.0011 & 0.0018
& -0.0024 & 0.0046
&  \\

\hline

\multicolumn{6}{l}{\textsl{Wald Test of Theoretical Planck Response ($H_0: a_{12} = 1.644$)}} \\
\multicolumn{1}{c}{} & \multicolumn{5}{l}{$\chi^2(1) = 0.32 \qquad \text{Prob} > \chi^2 = 0.57$} \\

\hline
\end{tabularx}
\begin{minipage}{\textwidth}\small
\textit{Notes:} 
Sensitivity is converted as $a_{21} \times 280$ ppm/K for comparison.
$^{***}$ denotes statistical significance at the 1\% level.
The Wald test evaluates the null hypothesis that the CTR equals the theoretical Planck response of 1.644 K.
\end{minipage}
\end{table} 

\begin{figure}[t!] 
\centering
\includegraphics[width=0.95\textwidth]{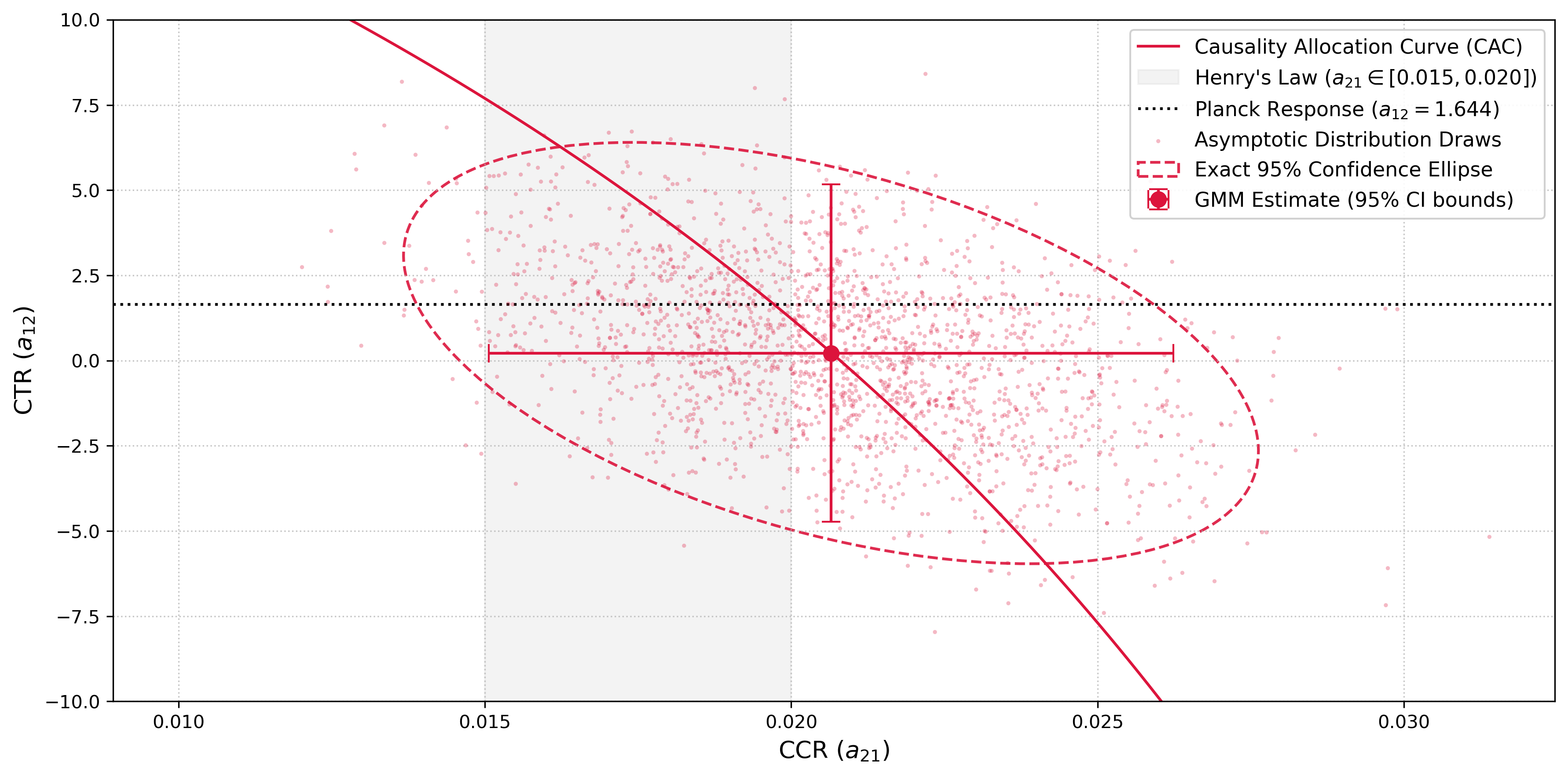}

\caption{Identification of structural parameters on the Causality Allocation Curve (CAC). The solid diagonal line represents the CAC derived from the residual covariance structure. The central marker and error bars indicate the GMM joint estimate (of Equations \eqref{eq:cac} and \eqref{eq:a21_reg}) and their individual 95\% confidence bounds. The dashed ellipse and scattered dots represent the exact 95\% joint confidence region and random draws from the asymptotic distribution, illustrating the structural trade-off (negative covariance) between the estimators. 
}
\label{fig:cac_ellipse}
\end{figure}

Figure \ref{fig:cac_ellipse} visualizes this structural identification. The plot displays the theoretical Causality Allocation Curve (CAC), representing the locus of $(a_{21}, a_{12})$ pairs satisfying the shock orthogonality condition. While the GMM point estimate for the CTR lies near the horizontal axis due to thermal inertia, its exact 95\% confidence ellipse fully encompasses the theoretical Planck response (horizontal dotted line). Simultaneously, the tight confidence interval for the CCR ($a_{21}$) perfectly aligns with the thermodynamic bounds of Henry's Law for oceanic outgassing (shaded vertical band). This visualizes a climate-carbon feedback architecture driven by an immediate temperature-to-carbon outgassing mechanism, while the reciprocal radiative forcing operates safely within its theoretical physical constraints.

\subsection{Structural Sensitivity and Response}

Based on the identification results in the previous section, we now evaluate the structural sensitivity of the climate--carbon system. 
First, to interpret the estimated reduced-form parameters ($\alpha, \beta, \Gamma$) in a structural context, we must consider the transformation $(I - A_0)$. 
While the cointegrating vector $\beta$ and the adjustment speeds $\alpha$ remain invariant to this transformation due to the property $\beta^\prime \Xi = 0$ (as formally derived in Appendix), the short-run lag coefficients $\Gamma$ must be structurally re-identified using the contemporaneous parameters $a_{12}$ and $a_{21}$.

By recalling the structural identity $(I - A_0) \Delta y_t = (I - A_0) \alpha \beta^\prime y_{t-1} + \Gamma^*_1 \Delta y_{t-1} + \dots + u_t$, the relationship between the structural and reduced-form coefficients is defined by the identity $\Gamma^*_1 = (I - A_0) \Gamma_1$. The specific element for the $\ln C_{t-1}$ to $T_t$ effect can be expanded as follows:
\begin{equation}
(\Gamma^*_1)_{12} = (\Gamma_1)_{12} - a_{12} (\Gamma_1)_{22} - a_{13} (\Gamma_1)_{32} \label{eq:gamma_struct}
\end{equation}
Owing to the strict exogeneity of orbital parameters, past climate states cannot influence current orbital forcing, which dictates that the reduced-form coefficient $(\Gamma_1)_{32}$ is effectively zero. 
Evaluating this expansion using our GMM point estimate ($\hat{a}_{12} = 0.2209$), the second term ($a_{12} (\Gamma_1)_{22} \approx 0.2209 \times 0.447$) becomes exceptionally small (approximately $0.1$). 
Consequently, the structural lag effect is overwhelmingly dominated by the first term, yielding $(\Gamma^*_1)_{12} \approx 15.1$.

Our structural estimation yields an Antarctic Earth System Sensitivity (ESS) of 13.0 K, which is remarkably consistent with the 11.1 K reported by \citet{Kaufmann2013}. This alignment is noteworthy given the methodological differences: while \citet{Kaufmann2013} utilized interpolated data and physical zero-constraints for identification, our results are derived from a gap-free sample with a data-driven GMM strategy. 
Specifically, the estimated long-term equilibrium for the 130 ka sample is given by $T - 18.748 \ln C = \text{const}$. This implies that a doubling of CO$_2$ is associated with a long-term Antarctic increase of $18.748 \times \ln(2) \approx 13.0$ K. Applying a polar amplification factor of approximately 0.5, this corresponds to a global ESS of $\sim$6.5 K, aligning with the reconstructed range of $9 \pm 4.5$ K by \citet{Snyder2016}.

Beyond the long-term equilibrium, the structural VECM reveals the specific causal architecture behind the ``slow adjustment'' of the climate system. While \citet{Kaufmann2016} identified adjustment scales of approximately 19 kyr, our isolation of the highly buffered CTR ($\hat{a}_{12} = 0.2209$) explains this delay as a fundamental manifestation of thermal inertia. Utilizing our structural point estimates, the cumulative temperature response within one millennium of a CO$_2$ doubling reaches 11.8 K in Antarctica (or 5.9 K globally). 

This response is driven by two mechanisms: the structural lag effect $(\Gamma^*_1)_{12} \times \ln(2) \approx 10.5$ K and the error correction process $\alpha \times (-13.0) \approx 1.3$ K. This result implies that while temperature triggers an immediate carbon release (CCR), the reciprocal radiative impact of CO$_2$ onto temperature is temporarily absorbed by the system's thermal inertia, manifesting primarily through these structural lags.

Finally, we compare these paleo-derived responses with modern climate observations. A 1,000-year response of 5.9 K global warming may appear high compared to the 1.2 K increase observed since the industrial revolution. However, this discrepancy is consistent with the non-linear ``front-loading'' of climate response: the immediate warming (Transient Climate Response) is followed by centuries of ``committed'' warming from slow internal feedbacks. The gap between the modern 1.2 K and our 1,000-year projection quantifies the massive ``thermal debt'' inherent in the Earth system's long-term dynamics.

Our GMM estimate $\hat{a}_{21} = 0.0206$ corresponds to a carbon cycle sensitivity of $\sim$5.77 ppm/K, 
which is broadly comparable in magnitude to
the $7.7$ ppm/K reported by \citet{Frank2010}. The stability of this CCR across different frameworks reinforces the warning by \citet{Kaufmann2016} that univariate analyses are prone to bias. By explicitly modeling the joint endogeneity, we ensure that our sensitivity parameters reflect the integrated dynamics of the climate--carbon system rather than isolated correlations.
\subsection{Impulse Response Simulation of Climate--Carbon Dynamics}

To visually clarify the feedback loops identified in our structural VECM, we perform impulse response function (IRF) simulations for two hallmark scenarios: a CO$_2$ doubling shock ($u_{2t} = \ln 2$) and a sudden 2 K temperature increase ($u_{1t} = 2.0$). 

\begin{figure}[t!]
\centering
\includegraphics[width=0.9\textwidth]{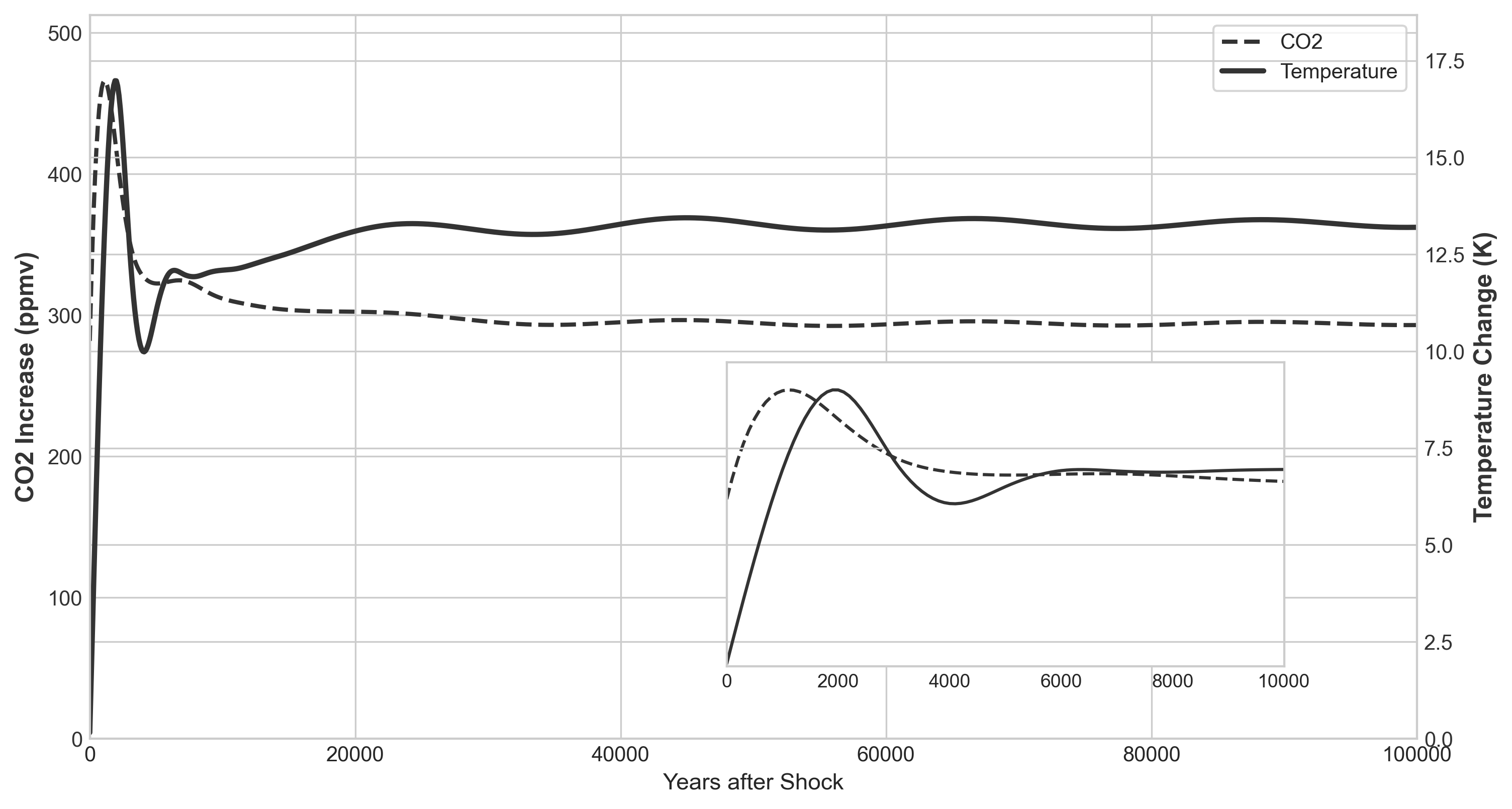}
\caption{Impulse response functions for a CO$_2$ doubling shock ($u_{2t} = \ln 2$). 
The plots show the evolution of CO$_2$ concentration and Antarctic temperature over 100,000 years and 10,000 years. 
Based on the GMM joint estimates, the initial temperature response is governed by the minute point estimate $\hat{a}_{12} = 0.2209$, 
which visually confirms the massive thermal inertia of the climate system. Consequently, the prominent warming does not occur instantaneously but manifests primarily through the structural lag effect within the subsequent millennium, gradually converging toward the long-term Earth System Sensitivity (ESS).}
\label{fig:irf_ln2}
\end{figure}

Figure \ref{fig:irf_ln2} illustrates the response to the CO$_2$ doubling shock. Governed by the GMM point estimate ($\hat{a}_{12} = 0.2209$), the immediate temperature response within the first 1,000-year step is heavily muted by thermal inertia. 
Instead, the prominent warming is triggered by structural lag effects, gradually converging toward the long-term ESS of 13.0 K in Antarctica. 
This dynamic visualization aligns with \citet{Rohling2012}, who emphasize that climate sensitivity is a function of the timescale, with the full magnitude of slow feedbacks becoming apparent only as the system approaches a new steady state.

\begin{figure}[t!]
\centering
\includegraphics[width=0.9\textwidth]{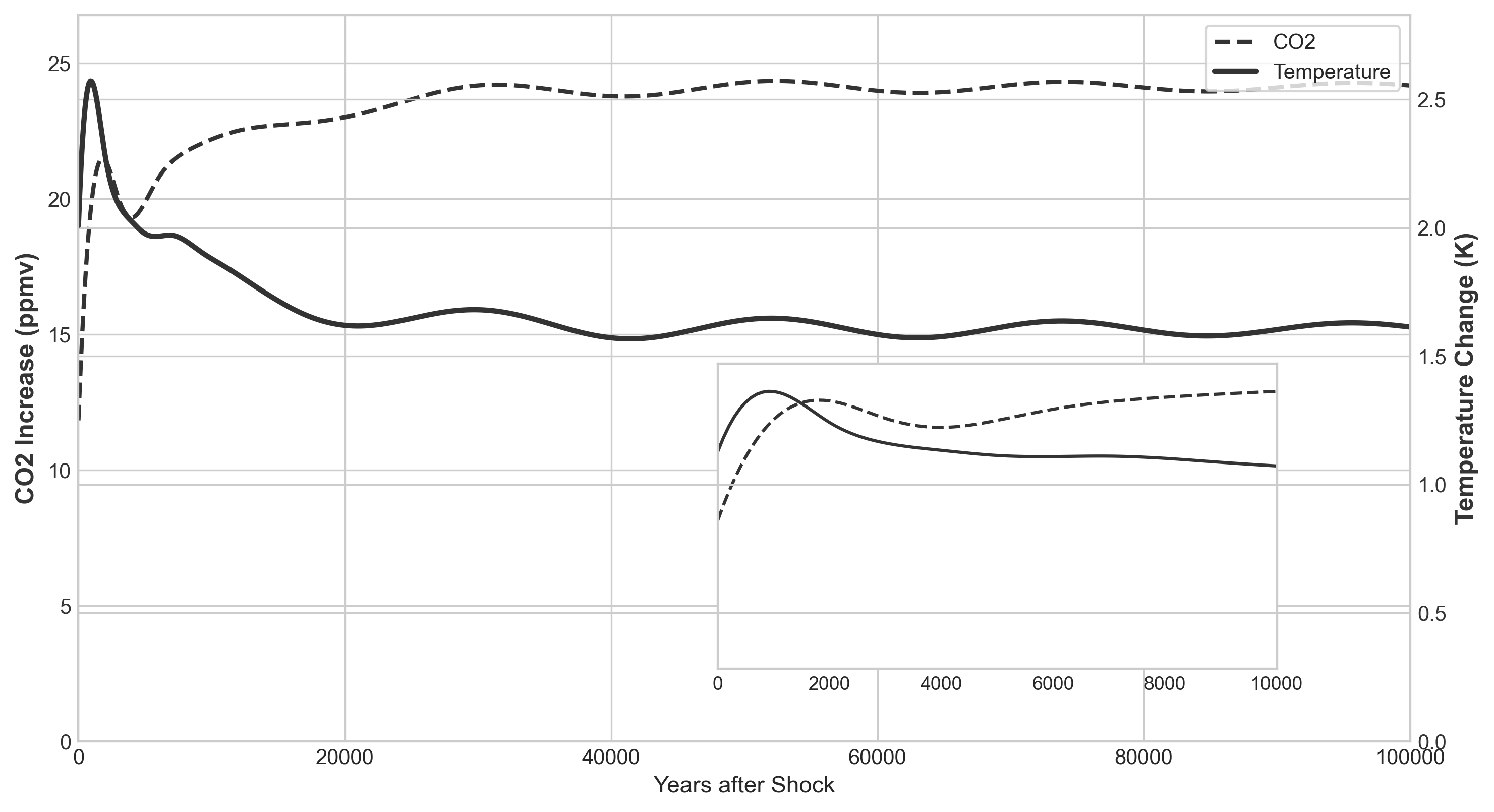}
\caption{Impulse response functions for a 2 K temperature shock ($u_{1t} = 2.0$). 
The simulation incorporates the statistically significant GMM estimate of the Contemporaneous Carbon Response ($\hat{a}_{21} = 0.0206$). The immediate and sharp rise in CO$_2$ visually captures this rapid outgassing mechanism consistent with Henry's Law, demonstrating that external thermal forcing instantaneously drives carbon fluctuations at the millennial scale before slower feedback loops propagate through the system.}
\label{fig:irf_2k}
\end{figure}

In contrast, Figure \ref{fig:irf_2k} displays the response to a 2 K temperature shock. 
Here, the carbon cycle responds instantaneously, reflecting the statistically significant CCR ($a_{21} > 0$). 
This immediate CO$_2$ release further amplifies the temperature anomaly through reciprocal lag effects, characterizing the self-reinforcing nature of the climate--carbon system. 
The distinct contrast between these two simulations provides clear visual evidence of the asymmetric contemporaneous feedback architecture at millennial resolutions.

\section{Concluding Remarks}

In this study, we have integrated rigorous time-series econometrics with paleoclimatic physics to re-examine the climate--carbon feedback architecture of the late Quaternary. 
By focusing on a gap-free 130,000-year interval and utilizing a structural VECM framework, we have circumvented the long-standing issues of interpolation bias and spurious regression that have often clouded the interpretation of ice-core records.

Our primary finding is the identification of a fundamental asymmetry in the contemporaneous climate--carbon interaction at the millennial scale. 
The structural results demonstrate that while temperature changes trigger an immediate and significant release of carbon from terrestrial and marine reservoirs (the CCR), the reciprocal radiative impact of CO$_2$ on temperature (the CTR) is heavily buffered by the Earth's massive thermal inertia within the initial 1,000-year time step. Crucially, however, the estimated confidence interval for this thermal response safely encompasses the theoretical Planck response, ensuring that the empirical model does not violate fundamental thermodynamic laws. 
This discovery provides a novel bridge between conflicting paleoclimatic narratives. 
The apparent ``synchronicity'' observed in recent deglacial chronologies can be understood as the rapid manifestation of the carbon outgassing (CCR), while the millennial-scale lag identified in earlier studies reflects the physical adjustment time (thermal debt) required for the greenhouse effect to alter the Earth's thermal state.

Furthermore, our model identifies a stable Earth System Sensitivity (ESS) of 6.5 K per CO$_2$ doubling. 
The fact that this estimate, derived through a dynamic structural framework, aligns so closely with previous OLS-based reconstructions suggests that the long-term coupling of the climate--carbon system is governed by a robust, persistent cointegrating trend that transcends simple glacial-interglacial oscillations.

Finally, this millennial-scale perspective offers a sobering context for modern climate change. 
Our analysis shows that the 1.2 K global warming observed since the industrial revolution is not only consistent with the sensitivity scale found in paleo-data but also represents merely the first stage of a much larger, non-linear adjustment process. 
The discrepancy between current observations and our 1,000-year 5.9 K projection quantifies a massive ``thermal debt'' inherent in the Earth system. 
As the system moves toward the long-term equilibrium identified in this study, the committed warming from centuries of cumulative slow feedbacks will likely present a challenge of unprecedented scale for the future of the planet.
\appendix
\normalsize
\renewcommand{\thesection}{\Alph{section}}
\makeatletter
\renewcommand{\@seccntformat}[1]{Appendix \csname the#1\endcsname\quad}
\makeatother
\numberwithin{equation}{section}
\section{The Impact of Data Interpolation on Structural Identification} \label{apd1}


To demonstrate the robustness of our methodological choice---specifically, the reliance on the gap-free 130 ka sample---we perform a supplementary analysis using the entire 800 ka ice-core record. Because the 800 ka dataset contains numerous observational gaps at the 1,000-year resolution, we applied linear interpolation to enforce a continuous grid before estimating the VECM. We then applied the identical GMM structural identification strategy (Equations \eqref{eq:cac} and \eqref{eq:a21_reg}) to this interpolated sample.

\begin{figure}[h!] 
\centering
\includegraphics[width=0.85\textwidth]{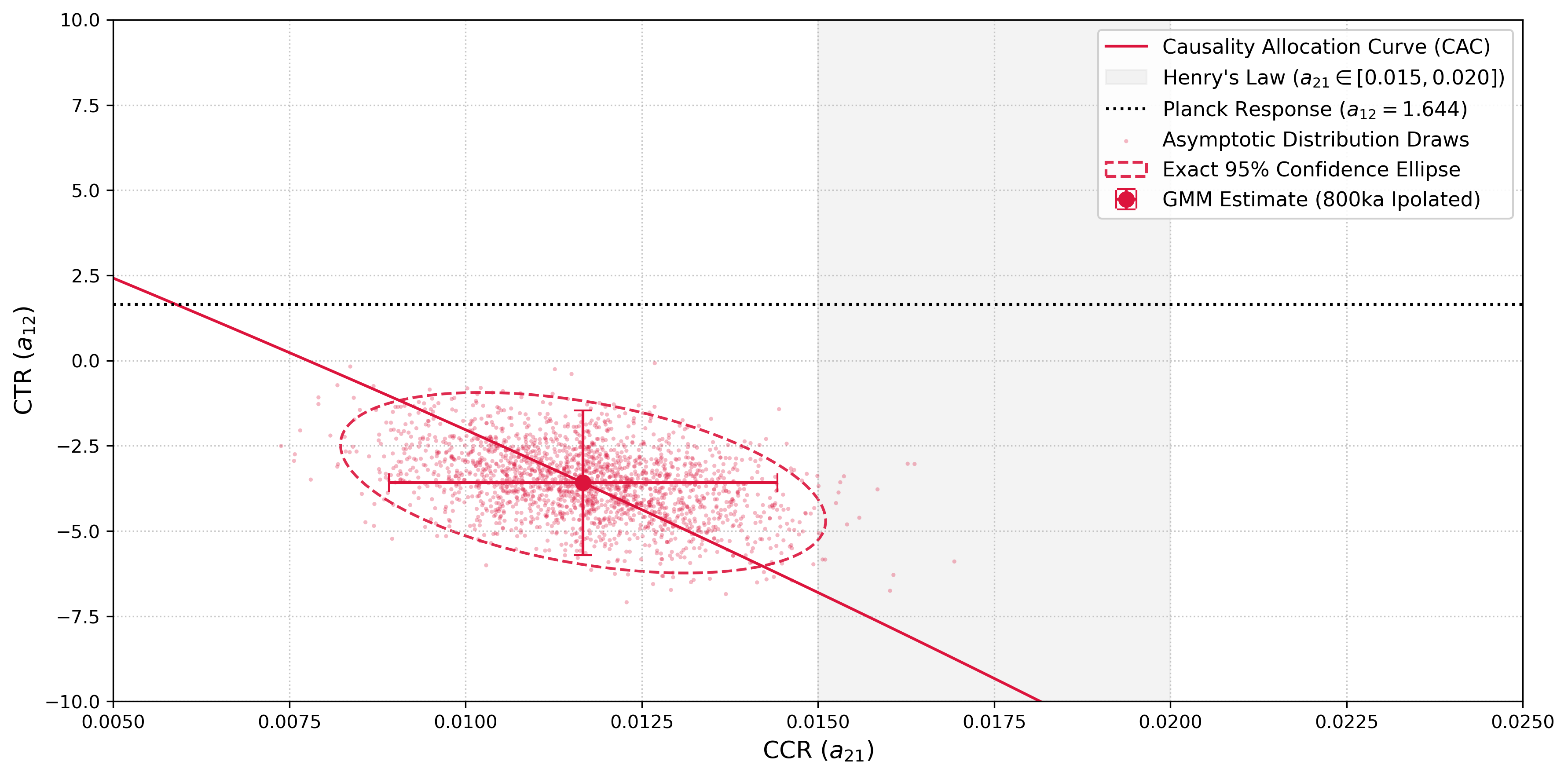}
\caption{Structural identification using the interpolated 800 ka sample. The estimated 95\% confidence ellipse is severely displaced, failing to align with the theoretical range of Henry's Law (shaded region) and sinking deeply into a physically impossible negative CTR territory, contradicting the theoretical Planck response (horizontal dotted line). The red dots represent random draws from the estimated asymptotic distribution.}
\label{fig:appendix_cac}
\end{figure}

The results yield an estimated contemporaneous carbon response (CCR) of $a_{21} \approx 0.0117$ and a contemporaneous temperature response (CTR) of $a_{12} \approx -3.58$ ($p < 0.01$). Figure \ref{fig:appendix_cac} visualizes these estimates on the Causality Allocation Curve (CAC) derived from the 800 ka sample residuals.

As visually evident in Figure \ref{fig:appendix_cac}, the inclusion of interpolated data causes the structural estimation to collapse into physically impossible territories. The CCR ($a_{21}$) falls significantly short of the thermodynamic limits dictated by Henry's Law. More critically, the CTR ($a_{12}$) becomes significantly negative, which would erroneously imply that an instantaneous increase in atmospheric CO$_2$ strongly cools the Antarctic climate---a clear violation of basic greenhouse gas physics and the Planck response.

This severe distortion is not a failure of the model, but rather a textbook manifestation of the statistical artifacts warned of by Miller (2019). Linear interpolation artificially smooths high-frequency variations, fundamentally altering the covariance structure of the reduced-form innovations ($s_{11}, s_{12}, s_{22}$). Because our structural identification relies precisely on the geometry of these residual covariances to allocate causality, the artificial smoothing bends the CAC, forcing the GMM estimator to land on a spurious intersection.

These supplementary findings strongly validate our empirical strategy. To accurately identify pure, contemporaneous physical feedbacks (such as outgassing and thermal inertia) without spurious bias, it is absolutely essential to utilize a sample that is entirely free of interpolation. Consequently, the gap-free 130 ka record (the ``Island'') utilized in the main text is not merely a subset of convenience, but the only mathematically and physically sound basis for this structural analysis.
\section{The VECM Representation and Impulse Responses} \label{apd2}

\subsubsection*{Decay of Shocks in a Stationary VAR}
Consider the following VAR($p$) model:
\begin{equation}
y_t = \Phi_1 y_{t-1} + \Phi_2 y_{t-2} + \dots + \Phi_p y_{t-p} + \epsilon_t
\label{VARp2}
\end{equation}
When the stability condition\footnote{All roots of the characteristic equation $\text{det}(I - \Phi_1 z - \dots - \Phi_p z^p)=0$ lie outside the unit circle in the complex plane. In this case, $z=1$ is not a root, and $\text{det}(G = I - \sum \Phi_i) \neq 0$ holds. Such a system is also referred to as being stationary.} is satisfied, this equation can be transformed into the following MA($\infty$) representation:
\begin{align}
y_t = \Psi_0 \epsilon_{t} + \Psi_1 \epsilon_{t-1} +\Psi_2 \epsilon_{t-2} + \dots
=
\sum_{h=0}^\infty \Psi_{h} \epsilon_{t-h} \quad (\Psi_0 = I)
\label{MAinfty}
\end{align}
Here, the coefficient matrix $\Psi_h$ represents the impact (impulse response) of a current shock $\epsilon_t$ on the level $y_{t+h}$ after $h$ periods, and it is defined recursively as $\Psi_h = \Phi_1 \Psi_{h-1} + \dots + \Phi_p \Psi_{h-p}$.

Assuming a constant shock has been applied from the past to the present ($\epsilon_{t-h} = \epsilon$), Equation \eqref{MAinfty} yields $y_t = (\sum_{h=0}^\infty \Psi_h) \epsilon$. On the other hand, assuming a long-run equilibrium state ($y_t = y_{t-1} = \dots = y$) and substituting this into Equation \eqref{VARp2}, we obtain $\left( I - \sum_{i=1}^p \Phi_i \right) y  = Gy = \epsilon$. By comparing the two, the following relationship is derived:
\begin{equation}
\sum_{h=0}^\infty \Psi_h 
= G^{-1}
\quad (\text{Long-run Multiplier})
\label{longrun}
\end{equation}
Under the stability condition, the right-hand side (long-run multiplier) is determined as a finite value, which requires the series on the left-hand side to converge to zero. As a necessary condition, the impulse response must vanish over time ($\lim_{h \to \infty} \Psi_h = 0$).

\subsubsection*{Redefinition via Differencing}
To encompass non-stationary cases, we redefine the level response $\Psi_h$ as the cumulative sum of responses in differences. When the difference $\Delta y_t = y_t - y_{t-1}$ is stationary, $\Delta y_t$ has the following MA representation:
\begin{equation}
\Delta y_t = C_0 \epsilon_t + C_1 \epsilon_{t-1} + C_2 \epsilon_{t-2} + \dots
=\sum_{j=0}^\infty C_j \epsilon_{t-j}
 \quad (C_0 = I)
 \label{diff}
\end{equation}
where $C_j = \Psi_{j} - \Psi_{j-1}$ represents the impact of the shock $\epsilon_{t-j}$ on the current ``change'' $\Delta y_t$. Given that the level $y_t$ is the cumulative sum of past differences $y_t = y_0 + \sum_{j=1}^t \Delta y_j$, the impulse response for the level, $\Psi_h = C_h + \Psi_{h-1}$, can be rewritten as:
\begin{align*}
\Psi_h = \sum_{j=0}^h C_j
\end{align*}

\subsubsection*{Application to Non-stationary Cases}
In systems where the characteristic equation has a unit root ($z=1$), the matrix $G$ becomes non-invertible ($\det(G)=0$), and the long-run multiplier in Equation \eqref{longrun} cannot be defined. Here, we treat the case where there is exactly one unit root, and all other roots lie outside the unit circle, as the non-stationary case. Such non-stationary series are referred to as $I(1)$, and their differenced series are stationary.

If the differenced series $\Delta y_t$ is stationary, then $\sum_{j=0}^\infty C_j$ in \eqref{diff} becomes a finite value. Therefore, $C_j$ must converge to zero ($\lim_{j \to \infty} C_j = 0$). Furthermore, in this case, $\Psi_h$ converges to a non-zero constant matrix $\Xi$:
\begin{equation}
\Psi_\infty = \lim_{h \to \infty} \Psi_h = \sum_{j=0}^\infty C_j = \Xi \neq 0
\label{xi_ref}
\end{equation}
This $\Xi$ refers to the permanent effect (Permanent Effect) of a temporary shock on the absolute level of the variables. In summary, while the impact of a shock in a stationary VAR is transitory and returns to zero, in a non-stationary system such as VECM, the impact on the "changes" accumulates and persists, leading to a permanent shift in the baseline level of the system.

The impulse response after $h$ periods can be decomposed as:
\begin{align*}
\Psi_h 
= \sum_{j=0}^h C_j
=  \sum_{j=0}^\infty C_j  -  \sum_{j=h+1}^\infty C_j 
= \Xi + \Upsilon_h 
\end{align*}
From Equation \eqref{xi_ref}, it is evident that $\lim_{h \to \infty} \Upsilon_h = \lim_{h \to \infty}(\Psi_h - \Xi) = 0$, meaning $\Upsilon_h$ represents the decaying transitory component.

\subsubsection*{Case with Cointegration: The Resilience of the Physical Bond}
The VAR($p$) model in Equation \eqref{VARp2} can be reformulated as the following VEC model:
\begin{align}
\Delta y_t = \Pi y_{t-1} + \Gamma_1 \Delta y_{t-1} + \dots + \Gamma_{p-1} \Delta y_{t-p+1} + \epsilon_t
\label{VEC}
\end{align}
Note that $\Phi_1 = I + \Pi + \Gamma_1$, $\Phi_i=\Gamma_i - \Gamma_{i-1}$, and $\Phi_p = -\Gamma_{p-1}$.
Given the non-stationarity of the level series $y_{t-1}$, $\Pi$ cannot be of full-rank.
The rank-dificient case $\Pi = \alpha \beta^\prime$ represents the long-run equilibrium (cointegrating) relationships. 
\citet{EG1987} demonstrated that for an $I(1)$ series $y_t$, a linear combination $\beta^\prime y_t$ exists that is stationary, since $\Delta y_t$ in Equation \eqref{VEC} must all be $I(0)$. 

A critical implication of cointegration is the orthogonality between the permanent effect $\Xi$ and the cointegrating vector $\beta$:
\begin{equation*}
\beta^\prime \Xi = 0
\end{equation*}
This property carries a profound physical meaning: while an exogenous shock (such as a sudden change in orbital forcing or a carbon cycle innovation) can permanently shift the absolute levels of temperature and {CO$_2$}, it is fundamentally incapable of breaking the equilibrium ``bond'' between them. 
The variables move to a new climatic baseline, but the underlying physical relationship, defined by $\beta$, remains intact. 

\citet{Johansen1991} provided the explicit representation for this permanent impact matrix:
\begin{align*}
\Xi = \beta_{\perp} (\alpha_{\perp}^\prime \Gamma \beta_{\perp})^{-1} \alpha_{\perp}^\prime
\end{align*}
where $\Gamma = I - \sum_{i=1}^{p-1} \Gamma_i$.
This expression confirms that the permanent shift $\Xi$ occurs only within the space spanned by $\beta_{\perp}$ (the orthogonal complement of $\beta$), ensuring that the long-run impact never deviates from the cointegrating attractor. 
Thus, the condition $\E[\beta^\prime y_t] = 0$ is maintained even as the Earth system undergoes permanent transitions between glacial and interglacial states.

\raggedright
\bibliography{references}

\subsubsection*{Data Availability}
The replication package, including the dataset and Stata/R codes used in this study, is available at \url{https://doi.org/10.7910/DVN/D5PQPT}.
\subsubsection*{Compliance with Ethical Standards}
The authors declare that they have no conflicts of interest.

\end{document}